\documentclass[11pt]{article}
\usepackage{amssymb}
\usepackage{amsmath}
\usepackage{amscd}
\usepackage{latexsym}

\catcode`@=11
\renewcommand{\section}{\@startsection{section}{1}{0pt}{\medskipamount}
{\medskipamount}{\large\bf}}
\numberwithin{equation}{section}
\catcode`@=12
\def\a{\alpha}
\def\b{\beta}
\def\g{\gamma}

\def\de{\delta}
\def\ve{\varepsilon}
\def\h{\eta}
\def\th{\theta}

\def\la{\lambda}
\def\m{\mu}
\def\n{\nu}
\def\r{\rho}
\def\s{\sigma}
\def\p{\phi}

\def\vk{\varkappa}

\def\o{\omega}
\def\Om{\Omega}
\def\La{\Lambda}

\def\1{\bar 1}
\def\2{\bar 2}
\def\3{\bar 3}
\def\4{\bar 4}

\def\hra{\hookrightarrow}
\newcommand{\unity}{\bf{1}}
\newcommand{\Qh}{\hat{Q}}
\newcommand{\Ah}{\hat{A}}
\newcommand{\ah}{\hat{a}}
\newcommand{\bh}{\hat{b}}
\newcommand{\gh}{\hat{g}}

\newcommand{\hth}{\hat{\theta}}
\newcommand{\yb}{\bar{y}}

\newcommand{\bb}{\bar{\beta}}

\newcommand{\zb}{\bar{z}}
\newcommand{\zeb}{\bar{\zeta}}

\newcommand{\ib}{\bar{\imath}}
\newcommand{\jb}{\bar{\jmath}}
\newcommand{\kb}{\bar{k}}
\newcommand{\lb}{\bar{l}}

\newcommand{\C}{\mathbb C}
\newcommand{\R}{\mathbb R}
\newcommand{\T}{\mathbb T}

\newcommand{\Zcal}{{\cal Z}}
\newcommand{\Acal}{{\cal A}}

\newcommand{\Fcal}{{\cal F}}
\newcommand{\Ecal}{{\cal E}}
\newcommand{\Mcal}{{\cal M}}
\newcommand{\J}{{\cal J}}
\newcommand{\U}{{\cal U}}

\newcommand{\with}{{\quad{\rm with}\quad}}

\def\im{\mbox{i}}
\def\N2{$N{=}2$}

\def\diff{\mbox{d}}
\def\tr{{\rm tr}}
\def\sfrac#1#2{{\textstyle\frac{#1}{#2}}}
\def\>{\rangle}
\def\<{\langle}
\def\+{\dagger}
\def\={\ =\ }
\def\und{\qquad\textrm{and}\qquad}
\def\and{\quad\textrm{and}\quad}
\def\for{\quad\textrm{for}\quad}

\begin{document}
\begin{titlepage}
\setcounter{page}{0}
\begin{flushright}
ITP--UH--03/13
\end{flushright}

\vskip 1.5cm

\begin{center}

{\Large\bf Instantons in six dimensions and twistors
}

\vspace{12mm}
{\large Tatiana A. Ivanova${}^*$, Olaf Lechtenfeld${}^{\+\times\circ}$, Alexander D. Popov${}^\+$
and Maike Torm\"ahlen${}^\+$}\\[8mm]

\noindent ${}^*${\em Bogoliubov Laboratory of Theoretical Physics, JINR\\
141980 Dubna, Moscow Region, Russia} \\
{Email: ita@theor.jinr.ru}\\[6mm]

\noindent ${}^\+${\em
Institut f\"ur Theoretische Physik\\
Leibniz Universit\"at Hannover \\
Appelstra\ss{}e 2, 30167 Hannover, Germany }\\
{Email: Olaf.Lechtenfeld, Alexander.Popov, Maike.Tormaehlen@itp.uni-hannover.de}\\[6mm]

\noindent ${}^{\times}${\em
Riemann Center for Geometry and Physics\\
Leibniz Universit\"at Hannover \\
Appelstra\ss{}e 2, 30167 Hannover, Germany }\\[6mm]

\noindent ${}^{\circ}${\em
Centre for Quantum Engineering and Space-Time Research\\
Leibniz Universit\"at Hannover \\
Welfengarten 1, 30167 Hannover, Germany }
\vspace{12mm}

\begin{abstract}
\noindent Recently, conformal field theories in six dimensions were discussed from the
twistorial point of view. In particular, it was demonstrated that the twistor transform
between chiral zero-rest-mass fields and cohomology classes on twistor space can be
generalized from four to six dimensions. On the other hand, the possibility of
generalizing the correspondence between instanton gauge fields and holomorphic bundles
over twistor space is questionable. It was shown by S\"amann and Wolf that holomorphic
line bundles over the canonical twistor space Tw($X$) (defined as a bundle of almost
complex structures over the six-dimensional manifold $X$) correspond to pure-gauge
Maxwell potentials, i.e. the twistor transform fails. On the example of $X=\C P^3$ we
show that there exists a twistor correspondence between Abelian or non-Abelian
Yang-Mills instantons on $\C P^3$ and holomorphic bundles over complex submanifolds of
Tw($\C P^3$), but it is not so efficient as in the four-dimensional case because the
twistor transform does not parametrize instantons by unconstrained holomorphic data as
it does in four dimensions.
\end{abstract}
\end{center}

\end{titlepage}

\section{Introduction and summary}

\noindent
Let us consider an oriented real four-manifold $X^4$ with a Riemannian metric
$g$ and the principal bundle $P(X^4, SO(4))$ of orthonormal frames
over $X^4$. The (metric) twistor space Tw$(X^4)$ of $X^4$ can be defined
as an associated bundle~\cite{AHS}
\begin{equation}\label{1.1}
\mbox{Tw}(X^4) = P\times_{{\rm SO}(4)} \mbox{SO(4)/U(2)}
\end{equation}
with the canonical projection Tw$(X^4)\to X^4$. This space parametrizes the almost complex
structures on $X^4$ compatible with the metric $g$ (almost Hermitian structures). It was
shown in~\cite{AHS, Penrose} that if the Weyl tensor of $(X^4, g)$ is anti-self-dual then
the almost complex structure  on the twistor space Tw$ (X^4)$ is integrable. Furthermore,
it was proven that the rank $r$ complex vector bundle $E$ over $X^4$ with an
anti-self-dual gauge potential $A$ over such $X^4$ lifts to a holomorphic bundle $\hat E$
over complex twistor space Tw$(X^4)$~\cite{AHS, Ward77}.

The essence of the canonical twistor approach is to establish a correspondence
between four-dimensional space $X^4$ (or its complex version) and complex twistor space
Tw$(X^4)$ of $X^4$.
Using this correspondence, one transfers data given on $X^4$ to data on
Tw$(X^4)$ and vice versa. In twistor theory one considers
{\it holomorphic\/} objects  $h$ on Tw$(X^4)$ (\v Cech cohomology
classes, holomorphic vector bundles etc.) and transforms them to objects
$f$ on $X^4$ which are constrained by some differential equations
\cite{AHS}-\cite{Wells}. Thus, the main idea
of twistor theory is to encode solutions of some differential equations
on $X^4$ in holomorphic data on the complex twistor space Tw$(X^4)$ of
$X^4$.

The twistor approach was recently extended to maximally supersymmetric Yang-Mills theory
on~$\C^6$~\cite{SWW}. It was also generalized to Abelian~\cite{SW1, MRT} and
non-Abelian~\cite{SW2} holomorphic principal 2-bundles over the twistor space
$Q_6\subset\C P^7\setminus \C P^3$, corresponding to self-dual Lie-algebra-valued 3-forms
on~$\C^6$. These forms are the most important objects needed for constructing (2,0)
superconformal field theories in six dimensions, which are believed to describe stacks of
M5-branes in the low-energy limit of M-theory~\cite{Moore}. Thus, it is worthwhile to
analyze the twistor transform in six dimensions in more detail.

We point out that there are some problems in generalizing the twistor approach to higher dimensions.
Namely, let $X^{2n}$ be a Riemannian manifold of dimension~$2n$. The metric twistor space of~$X^{2n}$
is defined as the bundle Tw$(X^{2n})\to X^{2n}$ of almost Hermitian structures on~$X^{2n}$
associated with the principal bundle of orthonormal frames of~$X^{2n}$, i.e.
\begin{equation}
\mbox{Tw}(X^{2n}):= P(X^{2n}, \mbox{SO}(2n))\times_{{\rm SO}(2n)}^{}\mbox{SO}(2n)/\mbox{U}(n)\ .
\end{equation}
It is well known that Tw$(X^{2n})$ can be endowed with an almost complex structure~$\J$,
which is integrable if and only if the Weyl tensor of~$X^{2n}$ vanishes when
$n>2$~\cite{BO}. This is a strong restriction on the geometry of $X^{2n}$ allowing only
conformally flat spaces, e.g. flat spaces and spheres, which may be not so interesting.
The restriction can be overcome if the manifold $X^{2n}$ has a $G$-structure (not necessary
integrable). In this case one can find a submanifold~$\Zcal$ of Tw$(X^{2n})$ associated
with the $G$-structure bundle $P(X^{2n},G)$ for $G\subset\ $SO($2n$), such that an
induced almost complex structure (also called~$\J$) on~$\Zcal$ is integrable. Many
examples were studied in~\cite{BO}-\cite{But}. Further problems can appear when
considering the twistor transform of holomorphic objects on Tw($X^{2n}$) or on
$\Zcal\hra\mbox{Tw}(X^{2n})$ to solutions of differential equations on $X^{2n}$. We will
discuss this for the case of $n=3$.

The papers~\cite{SW1, MRT} (see also references therein) show that twistor methods
can be useful in describing conformally invariant massless fields on the flat space
$\R^6\cong\C^3$ and its complexification $\C^6$ with the twistor space
\begin{equation}\label{1.3}
\mbox{Tw}(\R^{6})= Q_6\cong \R^{6}\times \C P^3\ .
\end{equation}
On the other hand, S\"amann and Wolf have shown~\cite{SW1} that holomorphic line bundles
over $\mbox{Tw}(\R^{6})$ trivial on all $\C P^3_x\hra \mbox{Tw}(\R^{6})$ correspond to
pure-gauge Maxwell potentials on $\R^6$, i.e. the twistor transform fails for the metric
twistor space $\mbox{Tw}(\R^{6})$. This was partially cured in~\cite{JMP12} where it was
shown that instantons on the six-sphere  $S^6=\R^{6}\cup\{\infty\}$ correspond to complex
vector bundles over the reduced twistor space $\hat\Zcal =G_2/$U$(2)\hra\mbox{Tw}(S^{6})$
with flat partial connections, where
\begin{equation}
\mbox{Tw}(S^6)= \mbox{Spin}(7)/\mbox{U}(3)
\end{equation}
is a compactification of the twistor space (\ref{1.3}). For the definition of the instanton equations
in dimensions higher than four and for some instanton solutions see e.g.~\cite{CDFN}-\cite{HN}.
Hence, constructing instanton configurations in six dimensions is a task more complicated than one might expect.

Instanton equations on the six-sphere $S^6$ are not quite standard since $S^6$ is a
nearly K\"ahler space with a nonintegrable almost complex structure. In fact, instantons
on $S^6$ are connections on {\it pseudo-holomorphic\/} bundles satisfying the
Donaldson-Uhlenbeck-Yau (DUY) equations~\cite{DUY}. Hence, for checking the power of the
twistor approach it is worthwhile to consider a K\"ahler 6-manifold. We choose the
complex projective space $\C P^3$ which can be considered as yet another compactification
of $\R^6\cong\C^3$.

On $\C P^3$ the DUY equations are the standard Hermitian Yang-Mills (HYM)
equations~\cite{DUY}. They are SU(3) invariant but not invariant under the SO(6)
Lorentz-type rotations of orthonormal frames. Therefore, one should describe them
with {\it reduced\/} twistor spaces. The DUY equations are well defined on
six-dimensional K\"ahler manifolds $X$ (as well as on nearly K\"ahler spaces~\cite{ChiSa,
Butr, Bryant3}), and their solutions are natural connections $\Acal$ on holomorphic
vector bundles $\Ecal\to X$~\cite{DUY}. As reduced twistor spaces of $\C P^3$ one can
consider
\begin{equation}\label{1.5}
\mbox{SU}(4)/\mbox{U}(2){\times}\mbox{U}(1)=:\Zcal\ \to\ \C P^3
\cong\mbox{SU}(4)/\mbox{U}(3)
\end{equation}
or
\begin{equation}\label{1.6}
\mbox{Sp}(2)/\mbox{U}(1){\times}\mbox{U}(1)=:\Zcal^\prime\ \to\ \C P^3
\cong\mbox{Sp}(2)/\mbox{Sp}(1){\times}\mbox{U}(1)
\end{equation}
which both are complex submanifolds of Tw($\C P^3$), with fibres $\C P^2$ and $\C P^1$,
respectively. We will show that bundles $(\Ecal , \Acal )$ over $\C P^3$ with HYM
connections $\Acal$ are pulled back to holomorphic vector bundles $(\tilde\Ecal ,
\tilde\Acal)$ over the reduced twistor spaces (\ref{1.5}) or (\ref{1.6}), depending on
the choice for $\C P^3$, being trivial along the fibres of the fibrations (\ref{1.5})
or~(\ref{1.6}), with a Hermitian Yang-Mills connection $\tilde\Acal$ on $\tilde\Ecal$.
Thus, contrary to the four-dimensional case, the twistor transform in six dimensions does
not parametrize instantons by unconstrained holomorphic data on the twistor space, since
the corresponding holomorphic bundles over $\Zcal$ and $\Zcal^\prime$ have to be
polystable. In other words, in four dimensions the twistor transform establishes a
correspondence between solutions of the instanton equations in $d=4$ and solutions of
holomorphic Chern-Simons theory on $d=6$ twistor space, but in six dimensions the twistor
transform establishes a correspondence between solutions of the instanton (HYM) equations
in $d=6$ and solutions of the HYM equations on the twistor space. The latter does not
facilitate solving the $d=6$ instanton equations. This is the outcome of our study of
instantons in six dimensions.

The structure of the remainder of this  paper is as follows. In Section 2 we portray the
space $\C P^3$ as a homogeneous space SU(4)$/$U(3) and Sp(2)$/$Sp(1)$\times$U(1), with
K\"ahler structures in both cases and allowing for the introduction of a quasi-K\"ahler
structure in the second case. In Section 3 we describe the geometry of the twistor spaces
$\Zcal$ and $\Zcal^\prime$ for SU(4)$/$U(3) and Sp(2)/Sp(1)$\times$U(1). In Section 4 we
study the twistor correspondence between instanton bundles over  $\C P^3$ and holomorphic
bundles over the proper twistor spaces.

\section{K\"ahler and quasi-K\"ahler structure on $\C P^3$}

\noindent
In this section we describe the geometry of the space $\C P^3$ as a homogeneous
manifold $\Mcal = {\rm Sp}(2)/{\rm Sp}(1){\times}{\rm U}(1)$ fibred over the four-sphere
$S^4$. We find it useful to describe orthonormal coframes on $S^4$, $S^2$
and $\Mcal$ in local coordinates. First, we choose a representative
element $Q\in\ $Sp(2) of the coset space $S^4{=}\ $Sp(2)/Sp(1)${\times}$Sp(1). Then,
expanding the flat connection $\Acal_0=Q^{-1}\diff Q$ into a basis of the Lie algebra
$sp(2)$, we obtain local (1,0)-forms $\th^1$ and $\th^2$ on an open subset $U$ of $S^4$ as
well as self-dual and anti-self-dual connections ($A^+$ resp.\ $A^-$) on Sp(1)-bundles 
over $S^4$. Using a representative element $g\in\ $SU(2) of the coset space
$S^2{=}\ $SU(2)/U(1)$\ \cong\ $Sp(1)/U(1), we get a local (1,0)-form $\th^3$ on $S^2$ and the
monopole connection $a$ on the Hopf bundle $S^3\to S^2$. After this, we combine $Q$ and
$g$ into a representative $\hat Q$ of the coset $\Mcal$ and arrive at local (1,0)-forms
$\hat \th^i$ on this coset, together with their Maurer-Cartan relations (\ref{2.25}).
Finally, changing an almost complex structure on $\Mcal$ via (\ref{2.29}), we find a
quasi-K\"ahler structure on the considered coset space.

\bigskip

\noindent {\bf Coset representation of $S^4$.\ } 
Let us consider  the group Sp(2)
fibred over $S^4=\,$Sp(2)/Sp(1)$\times$Sp(1),
\begin{equation}\label{2.1}
{\rm Sp}(2)\to S^4
\end{equation}
i.e. consider Sp(2) as the fibre bundle $P(S^4$, Sp(1)$\times$Sp(1)) with the structure
group Sp(1)$\times$Sp(1). Local sections of the fibrations (\ref{2.1}) can be chosen as
4$\times$4 matrices
\begin{equation}\label{2.2}
Q:= f^{-\frac{1}{2}}\begin{pmatrix}{\bf 1}_2 &- x\\x^\+ &
{\bf 1}_2\end{pmatrix}
\and
Q^{-1} =  Q^\+ =f^{-\frac{1}{2}}\begin{pmatrix}{\bf 1}_2 & x\\
-x^\+ & {\bf 1}_2\end{pmatrix}\in \mbox{Sp}(2)\subset\mbox{SU}(4)\ ,
\end{equation}
where
\begin{equation}\label{2.3}
x=x^\m\tau_\m\ ,\quad x^\+=x^\m\tau_\m^\+\ ,\quad
f:=1+x^\+ x= 1+r^2=1+\de_{\m\n}x^\m x^\n\ ,
\end{equation}
and the matrices
\begin{equation}\label{2.4}
(\tau_\m )= (-\im\s_i, {\bf 1}_2)\und
(\tau_\m^\+ )= (\im\s_i, {\bf 1}_2)
\end{equation}
obey
$$
\tau_\m^\+ \tau_\n = \de_{\m\n}\cdot {\bf 1}_2 + \eta_{\m\n}^i\,\im\,\s_i=:
\de_{\m\n}\cdot {\bf 1}_2 + \eta_{\m\n}\ ,\quad
\{\eta_{\m\n}^i\}{=}\{-\eta_{\n\m}^i\}{=}\{\ve^i_{jk}, \m{=}j, \n{=}k; \ \de_j^i,
\m{=}j, \n{=}4\}\ ,
$$
\begin{equation} \label{2.5}
\tau_\m\tau_\n^\+ {=} \de_{\m\n}\cdot {\bf 1}_2 {+} \bar\eta_{\m\n}^i\, \im\,\s_i{=:}
\de_{\m\n}\cdot {\bf 1}_2 + \bar\eta_{\m\n}\ ,\quad
\{\bar\eta_{\m\n}^i\}{=}\{-\bar\eta_{\n\m}^i\}{=}\{\ve^i_{jk}, \m{=}j, \n{=}k; \
-\de_j^i, \m{=}j, \n{=}4\} .
\end{equation}

\bigskip
\noindent Here $\{x^\m\}$ are local coordinates on an open set $\U\subset S^4$. The matrices
(\ref{2.2}) are representative elements for the coset space
$S^4=\,$Sp(2)/Sp(1)$\times$Sp(1).

\smallskip

\noindent
{\bf Flat connection on $S^4$.\ } Consider
a flat connection $\Acal_0$ on the trivial vector bundle $S^4\times\C^4\to S^4$
given by the one-form
\begin{equation}\label{2.6}
\Acal_0 = Q^{-1}\diff Q =:
\begin{pmatrix}A^- &-\p\\ \p^\+ & A^+\end{pmatrix} \ ,
\end{equation}
where from (\ref{2.2}) we obtain
\begin{equation}\label{2.7}
A^-= \sfrac{1}{f}\bar\h_{\m\n}x^\m\diff x^\n =:
\begin{pmatrix}\a_- &-\bar\b_-\\ \b_- & -\a_-\end{pmatrix}\in su(2) \ ,
\end{equation}
\begin{equation}\label{2.8}
A^+= \sfrac{1}{f}\h_{\m\n}x^\m\diff x^\n =:
\begin{pmatrix}\a_+ &-\bar\b_+\\ \b_+ & -\a_+\end{pmatrix}\in su(2) \ ,
\end{equation}
\begin{equation}\label{2.9}
\p = \frac{1}{f}\diff x = -\frac{\im}{f}
\begin{pmatrix}\diff x^3{+}\im\diff x^4&  \diff x^1{-}\im\diff x^2\\
\diff x^1{+}\im\diff x^2&  -(\diff x^3{-}\im\diff x^4)\end{pmatrix}
=-\frac{\im}{f}\begin{pmatrix}\diff z&\diff\bar y\\ \diff y&-\diff\bar z
\end{pmatrix}=:\begin{pmatrix}\th^2&\th^{\1}\\
-\th^1&\th^{\2}\end{pmatrix}\ ,
\end{equation}
with
\begin{equation}\label{2.10}
\a_+= \sfrac{1}{2f}(\yb\,\diff y + \zb\,\diff z - y\,\diff \yb -
z\,\diff \zb)\ ,\quad \b_+= \sfrac{1}{f}(y\,\diff z -z \,\diff y)\ ,
\end{equation}
\begin{equation}\label{2.11}
\a_-= \sfrac{1}{2f}(\yb\,\diff y + z\,\diff \zb - y\,\diff \yb -
\zb\,\diff z)\ ,\quad \b_-= \sfrac{1}{f}(y\,\diff \zb -\zb \,\diff y)\ ,
\end{equation}
\begin{equation}\label{2.12}
\th^1:=\frac{\im\diff y}{1+r^2}\ ,\quad
\th^2:=-\frac{\im\diff z}{1+r^2}\und
\th^{\1}:=-\frac{\im\diff \yb}{1+ r^2}\ ,\quad
\th^{\2}:=\frac{\im\diff \zb}{1+ r^2}\ .
\end{equation}
Here, the bar denotes complex conjugation.

\smallskip

\noindent
{\bf Coset representation of $S^2$.\ } Let us consider the Hopf bundle
\begin{equation}\label{2.13}
S^3\to S^2
\end{equation}
over the Riemann sphere $S^2\cong\C P^1$ and the one-monopole
connection $a$ on the bundle (\ref{2.13}) having in the local complex
coordinate $\zeta\in\C P^1$ the form
\begin{equation}\label{2.14}
a=\frac{1}{2(1+\zeta\bar\zeta)}\, (\bar\zeta\,\diff\zeta -\zeta\,
\diff\bar\zeta)\ .
\end{equation}
Consider a local section of the bundle (\ref{2.13}) given by the
matrix
\begin{equation}\label{2.15}
g=\frac{1}{(1+\zeta\bar\zeta)^{\frac{1}{2}}}\,
\begin{pmatrix}1&-\bar\zeta\\ \zeta & 1\end{pmatrix}
\in {\rm SU}(2)\cong S^3
\end{equation}
and introduce the $su(2)$-valued one-form (flat connection)
\begin{equation}\label{2.16}
g^{-1}\diff g=:
\begin{pmatrix}a&-\th^{\3}\\
\th^{3} & -a\end{pmatrix}
\end{equation}
on the bundle $S^2\times \C^2\to S^2$,
where
\begin{equation}\label{2.17}
\th^3=\frac{\diff\zeta}{1+\zeta\bar\zeta}\und
\th^{\3}=\frac{\diff\bar\zeta}{1+\zeta\bar\zeta}
\end{equation}
are the forms of type (1,0) and (0,1) on $\C P^1$ and $a$ is the
one-monopole gauge potential (\ref{2.14}).

\smallskip

\noindent {\bf Twistor space Tw($S^4$).\ }
Let us introduce 4$\times$4 matrices
\begin{equation}\label{2.18}
G=\begin{pmatrix}{\bf 1}_2&0\\0&g\end{pmatrix}\und\hat Q=QG\in
{\rm Sp}(2)\subset{\rm SU}(4)\ ,
\end{equation}
where $Q$ and $g$ are given in (\ref{2.2}) and (\ref{2.15}).
The matrix $\hat Q$ is a local section of the bundle
\begin{equation}\label{2.19}
{\rm Sp}(2) \to {\rm Sp}(2)/{\rm Sp}(1){\times}{\rm U}(1)=:\Mcal\ .
\end{equation}
Let us consider a trivial complex vector bundle ${\Mcal}{\times}\C^4\to\Mcal$
with the flat connection
\begin{equation}\label{2.20}
\hat\Acal_0 =\Qh^{-1}\diff\Qh=G^{-1}\Acal_0 G + G^{-1}\diff G =:
\begin{pmatrix} \Ah^-&-\hat\p\\ \hat\p^\+ & \Ah^+\end{pmatrix}\ ,
\end{equation}
where
\begin{equation}\label{2.21}
\hat\p = \p g=:\begin{pmatrix}\hat\th^2&\hat\th^{\1}\\
-\hat\th^1& \hat\th^{\2}\end{pmatrix}\ ,\quad
\Ah^-=A^-=\begin{pmatrix}\a_-&-\bb_-\\
\b_-& -\a_-\end{pmatrix}\ ,\quad
\Ah^+=:\begin{pmatrix}\hat\a_+&-\hat\th^{\3}\\
\hat\th^{3}& -\hat\a_+\end{pmatrix}\ ,
\end{equation}
with $\a_-$, $\b_-$ given in (\ref{2.11}) and
\begin{equation}\label{2.22}
\hat\a_+:=\frac{1}{1+\zeta\zeb}\left\{(1-\zeta\zeb)\,\a_+ + \zeb\b_+
-\zeta\bb_+ +\frac{1}{2}(\zeb\diff\zeta - \zeta\diff\zeb )\right\}\ ,
\end{equation}
\begin{equation}\label{2.23}
\hat\th^1:=\frac{1}{(1+\zeta\zeb)^{\frac{1}{2}}}\, (\th^1-\zeta\th^{\2})\ ,
\quad
\hat\th^2:=\frac{1}{(1+\zeta\zeb)^{\frac{1}{2}}}\, (\th^2+\zeta\th^{\1})\ ,
\end{equation}
\begin{equation}\label{2.24}
\hat\th^3:=\frac{1}{(1+\zeta\zeb)}\, (\diff\zeta +\b_+ -
2\zeta\a_+ +\zeta^2\bb_+)\ .
\end{equation}

{}From the flatness of the connection (\ref{2.20}), $\diff\hat\Acal_0 +
\hat\Acal_0\wedge\hat\Acal_0 =0$, we obtain the equations
\begin{equation}\label{2.25}
\diff\begin{pmatrix}\hat\th^1\\ \hat\th^2\\ \hat\th^3\end{pmatrix}+
\begin{pmatrix}-\hat\a_+-\a_-&\b_-&-\sfrac{1}{2R}\hat\th^{\2}\ \\[3pt]
-\bb_-&-\hat\a_++\a_-&\sfrac{1}{2R}\hat\th^{\1}\ \\[3pt]
\sfrac{R}{2\La^2}\,\hat\th^2& -\sfrac{R}{2\La^2}\,\hat\th^1& -2\hat\a_+\
\end{pmatrix}\wedge\begin{pmatrix}\hat\th^1\\ \hat\th^2\\ \hat\th^3\end{pmatrix}=0\ ,
\end{equation}
where we rescaled our one-forms $\hat\th$'s as
\begin{equation}\label{2.26}
\hat\th^1\to \frac{1}{2\La}\, \hat\th^1\ ,\quad \hat\th^2\to \frac{1}{2\La}\,\hat\th^2\und
\hat\th^3\to\frac{1}{2R}\,\hat\th^3\ .
\end{equation}
We see that (\ref{2.25}) defines the Levi-Civita connection with U(3) holonomy group
(K\"ahler structure) on $\Mcal$ if $R=\La$, where $R$ is the radius of $S^2$ and $\La$ is
the radius of $S^4$.

Note that the forms $\hat\th^i$ define on $\Mcal$ an
integrable almost complex structure $\J_+$~\cite{AHS} such that
\begin{equation}\label{2.27}
\J_+\,\hat\th^i = \im\,\hat\th^i
\end{equation}
with $i=1,2,3$. In other words, the $\hat\th^i$'s are (1,0)-forms with respect to $\J_+$
and the manifold $\Mcal$ with such a complex structure can be identified with the K\"ahler
manifold $\C P^3=\ $SU(4)/U(3) with the K\"ahler form
\begin{equation}\label{2.28}
\hat\omega :=\frac{\im}{2}\left (\hth^1\wedge\hth^{\1}+\hth^2\wedge\hth^{\2}+\hth^3\wedge\hth^{\3}\right )\ .
\end{equation}

\bigskip

\noindent
{\bf Quasi-K\"ahler structure on $\Mcal$.}
Recall that on the same manifold $\Mcal$ one can introduce the forms
\begin{equation}\label{2.29}
 \Theta^1:=\hat\th^1\ ,\quad \Theta^2:=\hat\th^2\und\Theta^3:=\hat\th^{\3}\ ,
\end{equation}
which are forms of type (1,0) with respect to an almost complex structure $\J_-$~\cite{ES},
$\J_-\,\Theta^i = \im\,\Theta^i$, which is a never integrable almost complex structure.
For $\Theta^i$ with the rescaling (\ref{2.26}) we have
\begin{equation}\label{2.30}
\diff\begin{pmatrix}\Theta^1\\ \Theta^2\\ \Theta^3\end{pmatrix} +
\begin{pmatrix}-\hat \a_+-\a_-&\b_-&0\\
-\bb_-&-\hat\a_++\a_-&0\\
0& 0& 2\hat\a_+
\end{pmatrix}
\wedge
\begin{pmatrix}\Theta^1\\ \Theta^2\\ \Theta^3\end{pmatrix}=
\frac{1}{2R}
\begin{pmatrix}\Theta^{\2}\wedge\Theta^{\3}\\ \Theta^{\3}\wedge\Theta^{\1}
\\ \frac{2R^2}{\La^2} \Theta^{\1}\wedge\Theta^{\2}\end{pmatrix}\ .
\end{equation}
The manifold $(\Mcal , \J_-)$ is a quasi-K\"ahler manifold. Recall that an almost
Hermitian $2n$-manifold with the fundamental (1,1)-form $\omega$ is called quasi-K\"ahler
if only (3,0)+(0,3) components of $\diff\omega$ are non-vanishing~\cite{Sal, Butr}. In
our case
\begin{equation}\label{2.32}
\o :=\sfrac{\im}{2}\bigl(\Theta^1\wedge\Theta^{\1}+\Theta^2\wedge\Theta^{\2}+
\Theta^3\wedge\Theta^{\3}\bigr)\ .
\end{equation}
 One can check that for arbitrary ratio
$\Lambda /R$ the (1,2) part of $\diff\omega$ vanishes and therefore $\Mcal$ is
quasi-K\"ahler \cite{ChiSa, ES}.

{}From (\ref{2.30}) one sees that the manifold $\Mcal=\ $Sp(2)/Sp(1)$\times$U(1) with an
almost complex structure $\J_-$ becomes a nearly K\"ahler manifold if $\La^2=2R^2$.
Recall that a six-manifold is called nearly K\"ahler if~\cite{Sal, ChiSa, Butr}
\begin{equation}\label{2.31}
\diff\o = 3\r\, \mbox{Im}\Om\for \Omega:=\Theta^1\wedge\Theta^2\wedge\Theta^3
\und \diff\Om = 2\rho\,\o\wedge\o \ ,
\end{equation}
where $\r\in\R$ is proportional to the inverse ``radius'' $\Lambda =\sqrt{2}R$ of $\Mcal$.

\bigskip

\section{Twistor spaces of $\C P^3$}

\noindent
Here we describe the geometry of the twistor spaces for the cosets SU(4)/U(3) and
Sp(2)/Sp(1)$\times$U(1) by using the same approach as in Section 2. First, we choose a
coset representative $V\in\ $SU(3) of $\C P^2=\ $SU(3)/U(2), introduce a coset
representative $\tilde Q = \hat Q\tilde V\in\ $SU(4) of SU(4)/U(2)$\times$U(1) and derive
the Maurer-Cartan relations (\ref{3.18}) for (1,0)-forms $\tilde\th^a$ on the twistor
space SU(4)/U(2)$\times$U(1) of  $\C P^3$. Then we do the same for the twistor space
Sp(2)/U(1)$\times$U(1) of the coset Sp(2)/Sp(1)$\times$U(1)$\,\cong\C P^3$. Namely, we
choose a representative $\breve Q$ of the coset Sp(2)/U(1)$\times$U(1), construct
(1,0)-forms $\breve\th^a$ on it via expanding the flat connection $\Acal_0' = \breve
Q^{-1}\diff \breve Q$ into an $sp(2)$-basis and finally derive the Maurer-Cartan equations
(\ref{3.35}) for $\breve\th^a$.

\bigskip

\noindent {\bf Coset representation of $\C P^2$. \ }
Let us consider the projection
\begin{equation}\label{3.1}
\mbox{SU}(3)\to
 \mbox{SU(3)}/\mbox{U(2)}=\C P^2 \ .
\end{equation}
One can choose as a coset representative of $\C P^2$ a local section of the bundle (\ref{3.1})
given by the matrix
\begin{equation}\label{3.2}
V\=\frac{1}{\g}\,\begin{pmatrix}1&Y^\+\\-Y&W\end{pmatrix}\ :=\
\frac{1}{\g}\,\begin{pmatrix}1&\bar\la^{\1}&\bar\la^{\2}\\-\la^1&W_{11}&W_{12}\\
-\la^2&W_{21}&W_{22}\end{pmatrix} \ \in\mbox{SU(3)}\ ,
\end{equation}
where
\begin{equation}\label{3.3}
\g^2:=1+Y^\+Y = 1+\la^1\bar\la^{\1}+\la^2\bar\la^{\2} \und
W=W^\+=\g\cdot{\unity}_2 -\frac{1}{\g +1}\,YY^\+\ .
\end{equation}
Here $\la^1$ and $\la^2$ are local complex coordinates on a patch of $\C P^2$.
{}From (\ref{3.2}) and (\ref{3.3}) it is easy to see that
\begin{equation}\label{3.4}
WY =Y\and W^2=\g^2-YY^\+\qquad\Leftrightarrow\qquad
V^\+V={\unity}_3=VV^\+\ .
\end{equation}

\smallskip

\noindent
{\bf Twistor space of SU(4)/U(3).\ } Consider the coset space
\begin{equation}\label{3.5}
\Zcal := \mbox{SU}(4)/\mbox{U}(2)\times \mbox{U}(1)
\end{equation}
and the projection
\begin{equation}\label{3.6}
\pi : \ \mbox{SU}(4)/\mbox{U}(2)\times \mbox{U}(1)\to\mbox{SU}(4)/\mbox{U}(3)\cong
\C P^3
\end{equation}
with fibres $\C P^2$. Using the group element (\ref{3.2}) to parametrize the typical $\C P^2$-fibre
in (\ref{3.6}), we introduce a flat connection $\tilde\Acal_0$ on the trivial bundle $\Zcal\times\C^4
\to\Zcal$ as
\begin{equation}\label{3.7}
\tilde\Acal_0 =\tilde Q^{-1}\diff\tilde Q =\tilde V^\+\hat\Acal_0\tilde V + \tilde V^\+\diff\,\tilde V\ ,
\end{equation}
where
\begin{equation}\label{3.8}
\tilde Q=\hat Q\tilde V\in\mbox{SU}(4)\and\tilde
V:=\begin{pmatrix}V&0\\0&1\end{pmatrix}\with V\in \mbox{SU}(3)\ .
\end{equation}

The flat connection $\hat\Acal_0$ is given in (\ref{2.20}) but here we write it as
\begin{equation}\label{3.9}
\hat\Acal_0 =\begin{pmatrix}\a_-&-\bar\b_-&-\hat\th^2&-\hat\th^{\bar 1}\\[2pt]
\b_-&-\a_-&\hat\th^1&-\hat\th^{\bar 2}\\[2pt]
\hat\th^{\bar 2}&-\hat\th^{\bar 1}&\hat\a_+&-\hat\th^{\bar 3}\\[2pt]
\hat\th^{1}&\hat\th^{2}&\hat\th^{3}&-\hat\a_+
\end{pmatrix}=:\begin{pmatrix}B&-T\\T^\+ &-\hat\a_+\end{pmatrix}\ ,
\end{equation}
where
\begin{equation}\label{3.10}
B =\begin{pmatrix}\a_-&-\bar\b_-&-\hat\th^2\\
\b_-&-\a_-&\hat\th^1&\\
\hat\th^{\bar 2}&-\hat\th^{\bar 1}&\hat\a_+
\end{pmatrix}\ ,\quad
T:=\begin{pmatrix}\hat\th^{\bar 1}\\\hat\th^{\bar 2}\\ \hat\th^{\bar 3}\end{pmatrix}\und
T^\+ = (\hat\th^1\ \hat\th^2\ \hat\th^3)\ .
\end{equation}
Using (\ref{3.7}), we obtain the connection
\begin{equation}\label{3.11}
\tilde\Acal_0 =\begin{pmatrix} V^\+BV+V^\+\diff\,V&-V^\+T\\T^\+V&-\hat\a_+\end{pmatrix} =:
\begin{pmatrix}\tilde B&-\tilde T\\ \tilde T^\+ &-\hat\a_+\end{pmatrix}\with \tilde T=
\begin{pmatrix}\tilde\th^{\bar 1}\\\tilde\th^{\bar 2}\\ \tilde\th^{\bar 3}\end{pmatrix}
\end{equation}
and for the curvature $\tilde\Fcal_0=\diff\tilde\Acal_0 + \tilde\Acal_0\wedge\tilde\Acal_0$ we get
\begin{equation}\label{3.12}
\tilde\Fcal_0 =\begin{pmatrix} \diff\tilde B+\tilde B\wedge\tilde B-\tilde T\wedge\tilde T^\+&
-\diff\tilde T-(\tilde B+ \hat\a_+\cdot{\unity_3})\wedge\tilde T\\[3pt]
\diff\tilde T^\++\tilde T^\+\wedge(\tilde B+\hat\a_+\cdot\unity_3)&-\diff\hat\a_+-\tilde
T^\+\wedge\tilde T
\end{pmatrix}\ .
\end{equation}
We have
\begin{equation}\label{3.13}
\tilde B=V^\+BV + V^\+\diff V=:\begin{pmatrix}\tilde\a_-&\Upsilon^\+\\-\Upsilon&\Sigma\end{pmatrix}
\end{equation}
with
\begin{equation}\label{3.14}
\Sigma=:\begin{pmatrix}\tilde a-\tilde\a_-&-\bar{b}\\b&-\tilde a+\hat{\a}_+\end{pmatrix}
\and \Upsilon =:\begin{pmatrix}\tilde\th^4\\\tilde\th^5\end{pmatrix}\ .
\end{equation}
{}Flatness $\tilde\Fcal_0 =0$ of the connection (\ref{3.11}) yields
\begin{equation}\label{3.15}
\diff\begin{pmatrix}\tilde\th^{1}\\\tilde\th^{2}\\ \tilde\th^{3}\end{pmatrix} +
\begin{pmatrix}-\tilde\a_--\hat\a_+&0&0\ \\[3pt]
0&-\tilde a+\tilde\a_--\hat\a_+&-b\ \\[3pt]
0& \bar b& \tilde a-2\hat\a_+
\end{pmatrix}\wedge\begin{pmatrix}\tilde\th^1\\ \tilde\th^2\\ \tilde\th^3\end{pmatrix}=
\begin{pmatrix}\tilde\th^{24}+\tilde\th^{35}\\ -\tilde\th^{1\4}\\ -\tilde\th^{1\bar 5}\end{pmatrix}\ .
\end{equation}
{} From
\begin{equation}\label{3.16}
\diff\tilde B + \tilde B\wedge\tilde B - \tilde T\wedge\tilde T^\+ =0
\end{equation}
it follows that
\begin{equation}\label{3.17}
\diff\begin{pmatrix}\tilde\th^{4}\\\tilde\th^{5}\end{pmatrix} +
\begin{pmatrix}\tilde a-2\tilde\a_-&-\bar{b}\\
{b}&-\tilde a+\hat\a_+-\tilde\a_-
\end{pmatrix}\wedge\begin{pmatrix}\tilde\th^4\\ \tilde\th^5\end{pmatrix}=
\begin{pmatrix}\tilde\th^{1\2 }\\ \tilde\th^{1\3}\end{pmatrix}\ .
\end{equation}
We obtain
\begin{equation}\label{3.18}
{\diff}\begin{pmatrix}\tilde\th^{1}\\\tilde\th^{2}\\ \tilde\th^{3}\\ \tilde\th^{4}\\
\tilde\th^{5}\end{pmatrix}{+}\begin{pmatrix}{-}\tilde\a_-{-}\hat\a_+&0&0&0&0\\
0&-\tilde a+\tilde\a_-{-}\hat\a_+&-b&0&0\\
0&\bar b&\tilde a-2\hat\a_+&0&0\\
0&0&0&\tilde a{-}2\tilde\a_-&-\bar{b}\\
0&0&0&b&{-}\tilde a{-}\tilde\a_-{+}\hat\a_+
\end{pmatrix}{\wedge}
\begin{pmatrix}\tilde\th^1\\ \tilde\th^2\\ \tilde\th^3\\ \tilde\th^4\\ \tilde\th^5\end{pmatrix}{=}
\begin{pmatrix}\tilde\th^{24}{+}\sfrac{\Lambda}{R}\tilde\th^{35}\\[2pt]
-\tilde\th^{1\4}\\[2pt]
-\sfrac{R}{\Lambda}\tilde\th^{1\bar 5}\\[2pt]
\sfrac{1}{4\Lambda^2}\tilde\th^{1\2}\\[2pt]
\sfrac{1}{4\Lambda R}\tilde\th^{1\3}\end{pmatrix} ,
\end{equation}
where we rescaled our $\tilde\th^a$ with $a=1,\ldots,5$ as in (\ref{2.26}):
\begin{equation}\label{3.19}
\tilde\th^1\to\frac{1}{2\La}\,\tilde\th^1\ ,\quad
\tilde\th^2\to\frac{1}{2\La}\,\tilde\th^2\ ,\quad
\tilde\th^3\to\frac{1}{2R}\,\tilde\th^3\ ,\quad
\tilde\th^4\to\tilde\th^4\und
\tilde\th^5\to\tilde\th^5\ .
\end{equation}
The manifold SU(4)/U(2)$\times$U(1) is the twistor space for the K\"ahler space $\C P^3=\ $SU(4)/U(3) for $\La^2=R^2$. Forms
$\tilde\th^a$ define on SU(4)/U(2)$\times$U(1) an integrable almost complex structure $\tilde\J_+$ such that
\begin{equation}\label{3.20}
\tilde\J_+\tilde\th^a = \im \, \tilde\th^a\ .
\end{equation}
In the K\"ahler case we choose $\Lambda =R=\sfrac12$.

\bigskip

\noindent{\bf Twistor space of Sp(2)/Sp(1)$\times$U(1).}
Consider the coset space
\begin{equation}\label{3.21}
 \Zcal' :=\mbox{Sp(2)}/\mbox{U(1)}\times\mbox{U(1)}
\end{equation}
and the projection
\begin{equation}\label{3.22}
 \pi' :\quad \mbox{Sp}(2)/\mbox{U(1)}\times\mbox{U(1)} \to \mbox{Sp}(2)/\mbox{Sp(1)}\times\mbox{U(1)}\cong\C P^3
\end{equation}
with fibres $\C P^1\cong\ $Sp(1)/U(1). We choose the group element
\begin{equation}\label{3.23}
\gh=\frac{1}{(1+\lambda\bar\lambda)^{\frac{1}{2}}}\,
\begin{pmatrix}1&-\bar\lambda\\ \lambda & 1\end{pmatrix}
\in {\rm SU}(2)\cong \mbox{Sp}(1)
\end{equation}
to parametrize the typical $\C P^1$-fibre in (\ref{3.22}), where $\lambda$ is a local
complex coordinate on the Riemann sphere $\C P^1$. By the formula
\begin{equation}\label{3.24}
\gh^{-1}\diff \gh=:
\begin{pmatrix}\ah&-\th^{\bar 4}\\
\th^{4} & -\ah\end{pmatrix}\ ,
\end{equation}
where
\begin{equation}\label{3.25}
\ah :=\frac{1}{2(1+\lambda\bar\lambda)}\, (\bar\lambda\,\diff\lambda -\lambda\,
\diff\bar\lambda )\ ,
\end{equation}
we introduce on $\C P^1$ the forms
\begin{equation}\label{3.26}
\th^4=\frac{\diff\lambda}{1+\lambda\bar\lambda}\und
\th^{\bar 4}=\frac{\diff\bar\lambda}{1+\lambda\bar\lambda}
\end{equation}
of type (1,0) and (0,1), respectively.

Using the group element (\ref{3.23}), we introduce a flat connection $\Acal_0'$ on the
trivial bundle $\Zcal'\times\C^4\to\Zcal'$ as
\begin{equation}\label{3.27}
\Acal_0' = \breve Q^{-1}\diff \breve Q =\hat G^\+\hat\Acal_0\hat G + \hat G^\+\diff\,\hat G\ ,
\end{equation}
where
\begin{equation}\label{3.28}
\breve Q=\hat Q\hat G\in\mbox{Sp}(2)\and\hat G:=\begin{pmatrix}\gh&0\\0&\unity_2\end{pmatrix}\in \mbox{Sp}(1)\subset
\mbox{Sp}(2)\ .
\end{equation}
The flat connection $\hat\Acal_0$ is given in (\ref{2.20}) and (\ref{3.9}). Using
(\ref{3.27}), we obtain the connection
\begin{equation}\label{3.29}
\Acal_0' =\begin{pmatrix} \gh^\+\Ah^-\gh+\gh^\+\diff\,\gh&-\gh^\+\hat\phi\\\hat\phi^\+\gh&\Ah^+\end{pmatrix} =:
\begin{pmatrix}\breve A^-&-\breve\phi\\  \breve\phi^\+ & \breve A^+\end{pmatrix}
\end{equation}
with
\begin{equation}\label{3.30}
\breve\phi=\gh^\+\hat\phi=\frac{1}{(1+\lambda\bar\lambda)^{1/2}}
\begin{pmatrix}\hat\th^2-\bar\lambda\hat\th^1&\hat\th^{\1}+\bar\lambda\hat\th^{\2}\\ -\hat\th^1-
\lambda\hat\th^2&\hat\th^{\2}-\lambda\hat\th^{\1}\end{pmatrix}
=:\begin{pmatrix}\breve\th^{2}&\breve\th^{\1}\\-\breve\th^{1}& \breve\th^{\2}
\end{pmatrix}\ ,
\end{equation}
\begin{equation}\label{3.31}
\breve A^+:=\begin{pmatrix}\breve\alpha_+&-\breve\th^{\3}\\\breve\th^{3}&- \breve\a_+\end{pmatrix}=
\begin{pmatrix}\hat\alpha_+&-\hat\th^{\3}\\\hat\th^{3}&- \hat\a_+\end{pmatrix}=\Ah^+\und
\breve A^-:=\begin{pmatrix}\breve\alpha_-&-\breve\th^{\bar 4}\\\breve\th^{4}&-
\breve\a_-\end{pmatrix}\ ,
\end{equation}
where
\begin{equation}\label{3.32}
\breve\a_-=\frac{1}{1+\la\bar\la}\,
\left\{(1-\la\bar\la){\a}_-+\bar\la{\b}_--\la\bar{{\b}}_-+\sfrac12\, (\bar\la\,\diff\la
-\la\diff\bar\la )\right\}\ ,
\end{equation}
\begin{equation}\label{3.33}
\breve\th^4=\frac{1}{1+\la\bar\la}\, \left\{\diff\la +{{\b}}_- -2\la{\a}_-
+\la^2\bar{\b}_-\right\}\ ,\quad \breve\th^{\4}:=\overline{\breve\th^4}\ .
\end{equation}

For the curvature $\Fcal'_0 =\diff\Acal'_0 +\Acal'_0\wedge\Acal'_0$ we get
\begin{equation}\label{3.34}
\Fcal'_0 =
\begin{pmatrix}
\diff\breve A^- +\breve A^-\wedge\breve A^- - \breve\phi\wedge\breve\phi^\+&-\diff\breve\phi -
\breve A^-\wedge\breve\phi - \breve\phi\wedge\breve A^+\\[2pt]
\diff\breve\phi^\+ +\breve \phi^\+\wedge\breve A^- + \breve A^+\wedge\breve \phi^\+&\diff\breve A^+ +
\breve A^+\wedge\breve A^+ - \breve\phi^\+\wedge\breve\phi
\end{pmatrix}\ .
\end{equation}
From the flatness $\Fcal'_0 =0$ of the connection  (\ref{3.29}) we obtain the
Maurer-Cartan equations
\begin{equation}\label{3.35}
\diff\,\begin{pmatrix}\breve\th^1\\\breve\th^2\\\breve\th^3\\\breve\th^4\end{pmatrix} +
\begin{pmatrix}-\breve\a_--\breve\a_+&0&0&0\\0&\breve\a_--\breve\a_+&0&0\\0&0&-2\breve\a_+&0\\0&0&0&-2\breve\a_-\end{pmatrix}\wedge
\begin{pmatrix}\breve\th^1\\\breve\th^2\\\breve\th^3\\\breve\th^4\end{pmatrix}=
\begin{pmatrix}-\breve\th^{2 4}-\breve\th^{3\2}\\\breve\th^{3\1}+\breve\th^{1\bar
4}\\2\breve\th^{12}\\-2\breve\th^{1\2}\end{pmatrix},
\end{equation}
which define the $u(1)\oplus u(1)$ torsionful connection on the twistor space $\Zcal' =\
$Sp(2)/U(1)$\times$U(1). The forms $\breve\th^a$ in (\ref{3.35}) with $a=1,\ldots,4$ define on
$\Zcal'$ an integrable almost complex structure $I_+'$ such that
\begin{equation}\label{3.36}
I_+' \breve\th^a = \im\breve\th^a\ .
\end{equation}
Its integrability follows from the vanishing (0,2)-type components of the torsion on the
right hand side of (\ref{3.35}).

\bigskip

\section{Twistor description of instanton bundles over $\C P^3$}

\noindent {\bf Instanton bundles over $\C P^3$.} Consider a complex vector bundle $\Ecal$
over $\C P^3$ with a connection one-form $\Acal$ having the curvature $\Fcal$. Recall
that ($\Ecal$, $\Acal$) is called an instanton bundle if $\Acal$ satisfies the Hermitian
Yang-Mills equations,\footnote{These equations are also called the Donaldson-Uhlenbeck-Yau equations.}
which on $\C P^3$  can be written in the form
\begin{equation}\label{4.1}
 \Fcal^{0,2}=0= \Fcal^{2,0}\qquad\Leftrightarrow\qquad\hat\Omega\wedge\Fcal =0\ ,
\end{equation}
\begin{equation}\label{4.2}
 \hat\o\lrcorner\,\Fcal=0\qquad\Leftrightarrow\qquad\hat\o\wedge\hat\o\wedge\Fcal =0\ ,
\end{equation}
where the notation $\hat\o\lrcorner$ exploits the underlying Riemannian metric
$g=\delta_{\ah\bh}e^{\ah}e^{\bh}$ on $\C P^3,\ \ah , \bh ,\ldots=1,\ldots,6$. Here, $\hat\o$
given in (\ref{2.28}) is a (1,1)-form, and $\hat\Om
:=\hat\th^1\wedge\hat\th^2\wedge\hat\th^3$ is a locally defined (3,0)-form on $\C P^3$.
Recall that, from the point of view of algebraic geometry, (\ref{4.1}) means
that the bundle $\Ecal\to\C P^3$ is holomorphic and (\ref{4.2}) means that  $\Ecal$ is a
polystable vector bundle~\cite{DUY}. In fact, in the right hand side of (\ref{4.2}) one
can add the term $\b\,\hat\o\wedge\hat\o\wedge\hat\o$ with $\b$ proportional to the first
Chern number $c_1(\Ecal )$, but we assume $c_1(\Ecal )=0$ since for a bundle with field
strength $\Fcal$ of non-zero degree one can obtain a degree-zero bundle by considering
$\check\Fcal=\Fcal - \sfrac1r\,(\tr\,\Fcal)\cdot\unity_r$, where $r=\,$rank\,$\Ecal$.

\bigskip

\noindent {\bf Pull-back to $\Zcal$.} Consider the twistor fibration (\ref{3.6}). Let
$(\widetilde\Ecal , \widetilde\Acal)=(\pi^*\Ecal , \pi^*\Acal)$ be the pulled-back
instanton bundle over $\Zcal$ with the curvature $\widetilde\Fcal =\diff\widetilde\Acal
+\widetilde\Acal\wedge\widetilde\Acal$. We have
\begin{equation}\label{4.3}
\widetilde\Fcal \=
\sfrac12\,\widetilde\Fcal_{ab}\,\tilde\th^{a}\wedge\tilde\th^{b}+\widetilde\Fcal_{a\bar b}\,
\tilde\th^{a}\wedge\tilde\th^{\bar b}+
\sfrac12\,\widetilde\Fcal_{\bar a\bar b}\,\tilde\th^{\bar a}\wedge\tilde\th^{\bar b}\=\pi^*\Fcal
\end{equation}
with $a,b,...=1,...,5$.
Using the relation between $\tilde\th^a$ and $\hat\theta^a$ described in Section 3, we
obtain
\begin{equation}\label{4.4}
\widetilde\Fcal_{\ib\jb}=C^{\kb}_{\ib}C^{\lb}_{\jb}\Fcal_{\kb\lb} \und
\widetilde\Fcal_{i\jb}=\bar C^{k}_{i}C^{\lb}_{\jb}\Fcal_{k\lb}\ ,
\end{equation}
where $C=\bar V^\+$ with
$$
C^{\1}_{\1}=\frac{1}{\g}\ ,\quad
C^{\1}_{\2}=-\frac{\la^1}{\g}\ ,\quad
C^{\1}_{\3}=-\frac{\la^{2}}{\g}\ ,
$$
\begin{equation}\label{4.10}
C^{\2}_{\1}=\frac{\bar\la^{\1}}{\g}\ ,\quad   C^{\2}_{\2}=\frac{\g
+1+\la^2\bar\la^{\2}}{\g(\g +1)}\ ,\quad C^{\2}_{\3}=-\frac{\la^2\bar\la^{\1}}{\g(\g +1)}\ ,
\end{equation}
$$
C^{\3}_{\1}=\frac{\bar\la^{\2}}{\g}\ ,\quad C^{\3}_{\2}=-\frac{\la^1\bar\la^{\2}}{\g(\g +1)}\ ,\quad
C^{\3}_{\3}=\frac{\g +1+\la^1\bar\la^{\1}}{\g(\g +1)}\ ,
$$
and $\bar C$ is the complex conjugate matrix.
Thus, more explicitly, we get
\begin{equation}\label{4.5}
\widetilde\Fcal_{\1\2}\=\frac{1}{\g}\, \left\{\frac{\g +1+\la^1\bar\la^{\1}}{\g +1} \Fcal_{\1\2} - \frac{\la^1\bar\la^{\2}}{\g+1}\Fcal_{\3\1}- \bar\la^{\2}\Fcal_{\2\3}\right\}\ ,
\end{equation}
\begin{equation}\label{4.6}
\widetilde\Fcal_{\3\1}\=\frac{1}{\g}\, \left\{\frac{\g +1+\la^2\bar\la^{\2}}{\g +1} \Fcal_{\3\1} -
\frac{\la^2\bar\la^{\1}}{\g +1}
\Fcal_{\1\2}- \bar\la^{\1}\Fcal_{\2\3}\right\}\ ,
\end{equation}
\begin{equation}\label{4.7}
\widetilde\Fcal_{\2\3}\=\frac{1}{\g}\, \left\{ \Fcal_{\2\3} + \la^1
\Fcal_{\3\1}+ \la^{2}\Fcal_{\1\2}\right\}\ ,
\end{equation}
\begin{equation}\label{4.8}
\widetilde\Fcal_{\ib\bar 4}\=\widetilde\Fcal_{\ib\bar 5}\=0\ ,
\end{equation}
\begin{equation}\label{4.9}
\widetilde\Fcal_{1\1}+\widetilde\Fcal_{2\2}+\widetilde\Fcal_{3\3}+\widetilde\Fcal_{4\bar 4}+
\widetilde\Fcal_{5\bar 5}=\Fcal_{1\1}+\Fcal_{2\2}+\Fcal_{3\3}\ .
\end{equation}

The vanishing of $\widetilde\Fcal_{\2\3}$ for all values of $(\la^1, \la^2)\in\C P^2$ is
equivalent to the holomorphicity equation (\ref{4.1}). In homogeneous coordinates
$y^i$ on $\C P^2$ ($\la^1=y^2/y^1,\ \la^2=y^3/y^1,\ y^1\ne 0$), this condition can be
written as
\begin{equation}\label{4.11}
\widetilde\Fcal_{\2\3}=0\quad\Leftrightarrow\quad y^i\ve_{ijk}\Fcal^{jk}=0\ ,
\end{equation}
where the indices $\ib , \jb , \ldots$ are raised with the metric $\delta^{i\jb}$. From
(\ref{4.5})-(\ref{4.9}) we see that solutions $\Acal$ of the HYM equations (\ref{4.1}),
(\ref{4.2}) on  $\C P^3$ correspond to solutions $\tilde\Acal=\pi^*\Acal$ of the HYM
equations on the twistor space $\Zcal$ of $\C P^3$, and $\tilde\Acal$ are flat connections
along fibres $\C P^2_x\hra\Zcal$. In other words, from (\ref{4.5})-(\ref{4.8}) we see
that the bundle $\widetilde\Ecal$ is holomorphic for holomorphic $\Ecal$ as well as
polystable due to (\ref{4.2}), (\ref{4.9}), and it is holomorphically trivial after
restricting to the fibres $\C P^2_x\hra\Zcal$ of the projection $\pi$ for each $x\in\C
P^3$. Vice versa, polystable holomorphic bundles over $\Zcal$  trivial on any fibre  $\C
P^2_x\hra\Zcal$ over $\C P^3$ correspond to solutions $\Acal$ of the HYM equations on $\C
P^3$. The only difference from the canonical twistor correspondence is that the bundle
$\tilde\Ecal$ is not only holomorphic~\footnote{
meaning it is defined by the equation $\bar{\partial}^2_{\tilde\Acal}=0$ of
holomorphic Chern-Simons theory for $\tilde\Acal$}
but also polystable, which is equivalent to imposing on $\tilde\Acal$ the additional
equation
\begin{equation}
\widetilde\Fcal_{1\1}+\widetilde\Fcal_{2\2}+\widetilde\Fcal_{3\3}+\widetilde\Fcal_{4\bar
4}+ \widetilde\Fcal_{5\bar 5}=0\ .
\end{equation}
Hence, the twistor transform does not help in solving the instanton equations in
six dimensions.

\bigskip

\noindent {\bf Pull-back to $\Zcal'$.} Consider now the twistor fibration (\ref{3.22})
and the pulled-back instanton bundle $(\Ecal', \Acal')=(\pi'^*\Ecal, \pi'^*\Acal)$ over
$\Zcal'$ with the curvature $\Fcal' =\diff\Acal' +\Acal'\wedge\Acal'$. We again have the
relation (\ref{4.3}) with $a,b,\ldots=1,\ldots,4.$ For the matrix $C$ in (\ref{4.4}) we now
find
\begin{equation}\label{4.12}
 C=\begin{pmatrix}\vk&\vk\la&0\\-\vk\bar\la&\vk&0\\0&0&1\end{pmatrix}\with
 \vk=(1+\la\bar\la)^{-\frac{1}{2}}\ ,
\end{equation}
where $\la$ is a local complex coordinate on $\C P^1$ used in (\ref{3.23})-(\ref{3.26}).

Using (\ref{4.12}), we obtain
\begin{equation}\label{4.13}
\Fcal'_{\1\2}= \Fcal_{\1\2}\ ,\quad\Fcal'_{\3\1}\=\vk (\Fcal_{\3\1} +
\bar\la\Fcal_{\2\3})\ , \quad\Fcal'_{\2\3}\=\vk (\Fcal_{\2\3} -\la\Fcal_{\3\1})\
,\quad \Fcal'_{\ib\bar 4}=0\ ,
\end{equation}
\begin{equation}\label{4.14}
\Fcal'_{1\1}+\Fcal'_{2\2}+\Fcal'_{3\3}+\Fcal'_{4\bar 4}=\Fcal_{1\1}+
\Fcal_{2\2}+\Fcal_{3\3}\ .
\end{equation}
Therefore, instanton bundles $(\Ecal, \Acal)$ over the nonsymmetric K\"ahler coset space
Sp(2)/Sp(1)${\times}$U(1)$\ \cong\C P^3$ are pulled back to holomorphic polystable
bundles $(\Ecal', \Acal')$ over the complex twistor space $\Zcal'=\
$Sp(2)/U(1)${\times}$U(1). Furthermore, $\Ecal'$ is flat along the fibres $\C P^1_x$ of
the bundle (\ref{3.22}), and one can set the components of $\Acal'$ along the fibres equal
to zero. Thus, the restrictions of the vector bundle  $\Ecal'$ to fibres $\C P^1_x\hra\Zcal'$
of the projection $\pi'$ are holomorphically trivial for each $x\in\
$Sp(2)/Sp(1)${\times}$U(1)$\ \cong\C P^3$. Note that (\ref{4.13}) and (\ref{4.14})
can be obtained from (\ref{4.5})-(\ref{4.9}) by putting $\la^1=-\la$ and $\la^2=0$.
Then (\ref{3.11}) will coincide with (\ref{3.29}) after the substitution $\tilde\th^{\4}\to -
\breve\th^{\4},\ \tilde\th^{\bar 5}\to -\breve\th^{2},\ b\to -\breve\th^{\1}$ etc.
This correspondence follows from the fact that $\Zcal'$ is a complex (codimension one) submanifold of the twistor space $\Zcal$.

\bigskip

\noindent
{\bf Acknowledgments}

\noindent
This work was partially supported by the Deutsche Forschungsgemeinschaft grant LE 838/13 and
the Heisenberg-Landau program.

\end{document}